\documentclass[sigconf, balance=false]{acmart}

\usepackage{graphicx} 
\usepackage{tabularx} 
\usepackage{ragged2e} 
\usepackage{color} 
\usepackage{anyfontsize} 
\usepackage[toc,page]{appendix} 
\usepackage{float} 
\usepackage{stfloats} 

\copyrightyear{2025}
\acmYear{2025}
\setcopyright{rightsretained}
\acmConference[CHI EA '25]{Extended Abstracts of the CHI Conference on
Human Factors in Computing Systems}{April 26-May 1, 2025}{Yokohama, Japan}
\acmBooktitle{Extended Abstracts of the CHI Conference on Human Factors in
Computing Systems (CHI EA '25), April 26-May 1, 2025, Yokohama,
Japan}\acmDOI{10.1145/3706599.3706674}
\acmISBN{979-8-4007-1395-8/25/04}

\begin{document}

\title{Creating benchmarkable components to measure the quality of AI-enhanced developer tools}

\author{Elise Paradis}
\email{eparadis@google.com}
\affiliation{%
  \institution{Google}
  \streetaddress{1265 Crossman Ave}
  \city{Mountain View}
  \state{California}
  \country{USA}
  \postcode{94089}
}

\author{Ambar Murillo}
\email{ambarm@google.com}
\affiliation{%
  \institution{Google}
  \streetaddress{Erika-Mann-Straße 33}
  \city{München}
  \country{Germany}
  \postcode{80636}
}

\author{Maulishree Pandey}
\email{pandeymauli@google.com}
\affiliation{%
  \institution{Google}
  \streetaddress{345 Spear St}
  \city{San Francisco}
  \state{California}
  \country{USA}
  \postcode{94105}
}

\author{Sarah D'Angelo}
\email{sdangelo@google.com}
\affiliation{%
  \institution{Google}
  \streetaddress{1600 Amphitheatre Pkwy}
  \city{Mountain View}
  \state{California}
  \country{USA}
  \postcode{94043}
}

\author{Matthew Hughes}
\email{matthew.hughes13@gmail.com}
\affiliation{%
  \institution{ServiceNow}
  \city{Santa Clara}
  \state{California}
  \country{USA}
}

\author{Andrew Macvean}
\email{amacvean@google.com}
\affiliation{%
  \institution{Google}
  \streetaddress{777 6th St South Building A}
  \city{Kirkland}
  \state{Washington}
  \country{USA}
  \postcode{98033}
}

\author{Ben Ferrari-Church}
\email{ferrarichurch@google.com}
\affiliation{%
  \institution{Google}
  \streetaddress{345 Spear St}
  \city{San Francisco}
  \state{California}
  \country{USA}
  \postcode{94105}
}

\renewcommand{\shortauthors}{Paradis et al.}

\begin{abstract}

In the AI community, benchmarks to evaluate \textit{model} quality are well established, but an equivalent approach to benchmarking products built upon generative AI models is still missing. This has had two consequences. First, it has made teams focus on \textit{model quality} over the developer experience, while successful products combine both. Second, product team have struggled to answer questions about their products in relation to their competitors.

In this case study, we share: (1) our process to create robust, enterprise-grade and modular components to support the benchmarking of the developer experience (DX) dimensions of our team's AI for code offerings, and (2) the components we have created to do so, including demographics and attitudes towards AI surveys, a benchmarkable task, and task and feature surveys. By doing so, we hope to lower the barrier to the DX benchmarking of genAI-enhanced code products.
\end{abstract}

\ccsdesc[500]{Human-centered computing~HCI design and evaluation methods}
\ccsdesc[500]{General and reference~Measurement}
\ccsdesc[500]{General and reference~Metrics}

\keywords{AI UX metrics; benchmarking; human computer interaction; evaluation; impact of AI}

\maketitle

\received{10 October 2024}

\section{Introduction}

As companies build new features and product experiences using generative artificial intelligence (henceforth, ``genAI''), they have to both execute on a novel vision \textit{and} evaluate the quality of what they are building. Doing so in a rigorous, benchmarkable way (i.e. one that is repeatable and comparable over time) is difficult for many reasons. First, teams have to agree on what to measure, create high-quality measurement instruments, and test them for quality. Second, agreeing on what task(s) to benchmark is non-trivial: what should be the difficulty of these tasks? Should there be one or many tasks? And third: genAI-enhanced products and features are ever-changing and doing so at breakneck pace. Merely a year ago, a simple prompt was novel to end users; today, those who have invested in their AI skills are creating agents to get tasks done on their behalf~\cite{wang2024}.

The main aim of this project was to support our team with high-quality data to evaluate the user experience of our AI for code features, including an estimate of the impact of AI-enhanced coding tools on developer velocity (see \cite{paradis2024rct}). This case study aims to answer the following, narrower questions: How might we support the creation of robust, enterprise-grade, and modular evaluation components to better understand the impact of AI on developer workflows, while systematically comparing the quality of features across products and over time? In other words: how might we create benchmarkable components that accelerate other researchers at Google and beyond in assessing the quality and impact of their AI features for code?

In this case study, we share how we created a set of robust, enterprise-grade and modular components to support the benchmarking of the user experience dimensions of Google's AI for code offerings. First, we review what exactly are benchmarks and argue for the need to mature the benchmarking space for the developer experience (DX). We also briefly review recent efforts to evaluate the impact of genAI products on the developer experience. We then turn to the steps we took to create our different benchmarkable components: from the identification of benchmarkable metrics to the crafting of questions evaluating developers' attitudes towards generative AI; from cognitive testing to task identification and piloting of all components to ensure feasibility. We move on to describe the final experimental design we used to bring all components together and evaluate the impact of generative AI coding tools on developer end-to-end velocity.\footnote{We defined velocity as the time taken to complete an enterprise-grade engineering task, with or without AI, from the branching of code all the way through to testing and building the code. Measuring the change in velocity attributable to AI was therefore the first and critical aim of this project. The results can be found on arXiv \cite{paradis2024rct}} We wrap up by sharing the impact our benchmark has had at Google and laying a vision for how others could either use the components we created or create their own.

We make the following contributions:

\begin{itemize}
    \item We describe, step by step, a process for setting up a rigorous benchmarking program to evaluate genAI product experiences (see \S\ref{sec:process})
    \item We share our research design to benchmark genAI features for code (see \S\ref{sec:researchDesign})
    \item We share the task stimulus, validated surveys, and study protocol for researchers and practitioners to use in their own benchmarking efforts (see Appendices \ref{app:taskDescr} and \ref{app:demogAttQ})
    \item We discuss the impact we have had and how different teams could use our components (see \S\ref{sec:impactnext})
\end{itemize}
\section{Background and Related Work}

\subsection{Current approaches to genAI benchmarking}
A benchmark is ``a standard or a point of reference against which things may be compared or assessed''~\cite{OED:Benchmark}, including products, people, or processes. Since the rise of large language models (henceforth, ``LLMs'') and genAI, companies have sought to benchmark the capabilities of their models against those of their competitors using standardized sets of tasks or datasets~\cite{mcintosh2024}. In the more specific context of product or feature benchmarks, we take benchmarking to mean the tasks, metrics, and research designs to evaluate the quality of an AI-enhanced coding tool across multiple dimensions such as usefulness, usability, and actual productivity gained.

In the AI-enhanced developer tools space as elsewhere in the industry, the focus has tended to be on the benchmarking of \textit{model} quality using automatic evaluation methods~\cite{mcintosh2024}, over and above the benchmarking of the user experience of the tools created \textit{based on} LLMs for code. For instance, ``academic'' benchmarks---such as BIG-bench~\cite{srivastava2023} or the Python-language PyBench~\cite{pybench}---are used broadly. There are currently no industry standards for measuring the user experience of AI-enhanced products or features for code. A cursory search of the Papers with Code website\footnote{\url{https://paperswithcode.com/}} for ``AI benchmarks'' yields 42 papers that cover quantum a broad range of topics, but \textit{none} related to the developer experience of genAI~\cite{paperswithcode}; a Google Scholar search shows a similar gap. 

This is likely due in part to the relative historical recency of productionized genAI. Indeed, the famous ``Attention is all you need'' paper introduced transformers to the world in 2017~\cite{waswani2017}, but ChatGPT and the ``chatbot revolution''~\cite{cnn-2023-chatRevolution} were launched only in November 2022, two years ago. In the expansionary period that has followed for AI products, it is not surprising that DX researchers simply have not had time to both support emerging products while also evaluating them robustly.

The absence of DX-focused standards or metrics against which to benchmark developer products has had two consequences. On the one hand, it has meant that product teams (including product managers, designers, and researchers) have struggled to answer questions about the quality of their AI products, how well they stack up to the competition, and how they are evolving over time based on specific standards. On the other, it has led to an over-steering towards model quality and away from the developer experience.

\subsection{Recent evaluations of genAI and the developer experience} \label{sec:recentevals}
To date, only a few evaluations of commercially available AI-en\-hanced developer tools have been published, and most of them have focused on GitHub Copilot, one of the first and most-widely used coding assistants available today. The GitHub team itself published two papers on the impact of the AI coding assistant, both of them controlled trials~\cite{peng2023, githubAcc}. The first study found a velocity gain of 56\% for developers using GitHub Copilot on a complex coding task---compared to developers \textit{not} using it---; it also found that 85\% of developers felt more confident in the quality of the code they produced with Copilot~\cite{peng2023}. The second study describes a collaboration with Accenture, a consulting firm, where developers were randomly assigned to be given access to Copilot or not~\cite{githubAcc}. The team found high adoption metrics and rates of acceptance of code suggestions, and reported highly positive rates of self-reported usage, ease of use, usefulness, use for production code, and fulfillment for Copilot users.

Our own estimate for the impact of three IDE-based genAI features for code at Google, obtained through the randomized controlled trial discussed here and using the instruments we share here, suggests that developers are 21\% faster with AI, controlling for factors such as average hours spent coding daily, seniority, and  previous usage of AI tools \cite{paradis2024rct}. This paper also suggested that those who code more and who are more senior in their careers are even faster with AI.

Other evaluations suggest that the story of the impact of AI coding tools on code quality or the developer experience might not be as straightforwardly positive. GitClear---a company that owns a data visualization product that evaluates code repositories for health and overall development trends---published a white paper suggesting that AI-enhanced tools might be exerting a downward pressure on code quality at the ecosystem level~\cite{gitclear2023}. The main causal pathway they identified suggests that less-experienced developers might accept and then productionize lower-quality code generated by AI, compared to code written by humans. 

Results seem similar at a more micro level. For instance, in a controlled experiment with 21 developers, Imai found that the use of Copilot as a pair programmer helped increase the number of lines of code added but also increased the number of lines of code that were subsequently deleted, compared to when humans acted as pair programmers~\cite{imai2022}. This further substantiates that AI might be generating lower-quality code. Vaithilingam\textit{ et al.}'s study with 24 developers found that participants enjoyed using the assistant, but that Copilot did \textit{not} improve task completion rates or success rates~\cite{vaithilingam2022}. They posit that issues with code understanding, editing and debugging plague the usefulness of AI-enhanced coding assistants~\cite{vaithilingam2022}. Finally, when it comes to usability, Zhang \textit{et al}. demonstrated that developers find integration of code suggestions by Copilot very difficult outside of VS Code~\cite{Zhang2023}. The authors concluded that use of Copilot might be a ``double-edged sword'' at this point in its history, given the few languages it supports well, difficulties with access, threats to code privacy, etc.~\cite{Zhang2023}. These mixed findings invite further research and analysis of the quality and impact of genAI-enhanced coding tools and point to the need for greater standardization within the field: to be able to compare across studies and across code assistants, we need to replicate studies; to do so, we need broadly available and reusable measurement instruments.

To conclude, combining a high-quality model \textit{and} a great user experience in a multi-faceted approach will be the way forward with genAI-enhanced products~\cite{AIsoftEng}. However, to date, much of the emphasis has been placed on model capabilities. Bringing a DX-centered lens to benchmarking genAI products promises to help change the focus, away from merely computational questions and towards the experiential components of the technology. How to safeguard developer productivity and happiness as developer tools change is where a robust approach to product benchmarking of genAI-enhanced products can be optimally helpful, and this is what the rest of this paper is dedicated to help solving.
\section{Creating a product benchmark}
\label{sec:process}

At this early stage in the development of AI-enhanced developer products, it is critical to evaluate their quality and impact on users, and to do so in a manner that enables comparison between products and over time. To enable a rigorous benchmarking approach, we had to answer a series of questions, such as ``What concepts should we benchmark?'', ``What are the factors that might influence developers' likelihood to complete a task and speed of task completion?'', ``What task or tasks should we benchmark?'', and ``What is the quality of our instruments, and is the planned study even feasible?'' The following sections answer these questions and documents how our team created the different components that came together in the experiment we ran to benchmark our products. You can learn more about our experimental design in Section~\ref{sec:researchDesign}.

\subsection{Step 1: Identification of benchmarkable concepts and survey question drafting}

To answer questions about which metrics to choose, we started from Google's internal metrics framework for the developer experience, which is a variant of the SPACE framework~\cite{forsgren2021} that includes more system efficiency concepts. We decided to center on sentiment and productivity metrics, and narrowed down the list to the metrics that were most applicable in the context of our study (see Table \ref{tab:metricsDefs}). 

\begin{table*}
\renewcommand{\arraystretch}{1.1}
\fontsize{8pt}{8pt}\selectfont
\centering
\caption{Selected metrics and their definitions}
\label{tab:metricsDefs}
\begin{tabular}{ l l l }
\hline
\textbf{Metrics} & \textbf{Submetrics} & \textbf{Definitions}\\
\hline
\textit{Sentiment} & & \\
\hline
Satisfaction &  & Overall satisfaction with AI feature\\
Trust in accuracy &  & The extent to which developers expect that the AI\\
 & & feature's outputs will be accurate \\ 
Goal importance &  & How central the task is to their work \\ 
Helpfulness &  & How helpful the AI feature is \\ 
Feature completeness &  & How polished the AI feature feels \\ 
\hline
\textit{Productivity} & & \\
\hline
Perceived code quality & &  \\
 & Quality of code suggestion & Quality of the output of the feature \\
 & Long-term impact on the code base & Expected long-term impact of feature\\
Ease of use & & How easy or difficult the feature is to use\\
Perceived productivity & & Expected productivity impact of feature use\\
\hline
\end{tabular}
\end{table*}

We then turned to the question of factors that would influence developers' time on task. We were interested in the factors that might impact developers' performance on a task, with or without access to AI-enhanced coding features. A recent systematic review pointed to different factors influencing AI adoption~\cite{kelly2023}, and pointed to perceived usefulness, expectations of performance and effort required to use, as well as attitudes towards, and trust in, AI. We grouped factors into three categories: demographic factors, current use of AI, and attitudes towards AI.

Among the demographic factors that could influence velocity on task, we selected the following concepts:

\begin{itemize}
\item Programming experience (years)
\item Tenure at Google (years)
\item Hours spent programming per day (hours)
\item Familiarity with language used for benchmarking task (C++)
\end{itemize}

For current use of AI, we selected the following concepts:
\begin{itemize}
\item Experience with AI tools
\item Expertise with AI overall
\end{itemize}

Once all factors were identified, we selected previously-validated survey questions from an internal repository. For factors where no relevant question existed internally, we conducted a review of the literature. We share findings from this literature review in the next section and in Appendix~\ref{app:litreview}, and our final list of questions in Appendices~\ref{app:demogAttQ} and \ref{app:taskFeaturesSurvey}.

\subsection{Step 2: Review of the literature on attitudes towards generative AI and question drafting}
As researchers, we stand on the shoulders of giants who have explored a topic before we have. The space of attitudes towards AI is no different, despite the recent emergence of generative AI. We therefore started our exploration of attitudes towards AI with a literature review. We reviewed both industry publications (such as survey reports from Jetbrains~\cite{jetbrains}, Stack Overflow~\cite{stacko}, and CoderPad~\cite{coderpad}), our internal Engineering Satisfaction Survey~\cite{dangeloEngSat}, and the academic literature on attitudes towards AI and technology more broadly~\cite{BERGDAHL2023,edison2003,SCHEPMAN2020}. We looked for published questions and scales that we could use and potentially compare with our study findings. More details can be found in Appendix~\ref{app:litreview}.

\subsection{Step 3: Cognitive testing}\label{sec:cogtest}

We engaged in two rounds of cognitive testing for our survey questions, following the ``think aloud'' paradigm identified by Beatty and Willis~\cite{beatty2007}. Cognitive testing (alternately called ``cognitive interviewing'') helps researchers ensure that participants understand the questions, can retrieve information from memory, can map their own answers to the answer scale, and can comfortably report an answer.

\subsubsection{First round: Task and feature questions.} 
The first round of cognitive testing was focused on the task and feature questions, and was conducted with 9 full-time software engineers at Google. We selected both native and non-native speakers of English. This was done to ensure the questionnaires would be understood by people with varied levels of English language ability, given the globally-distributed population of engineers our tools support.

After being shown the questions, the participants were asked to:

\begin{itemize}
\item Read the question out loud
\item Reformulate the question in their own words
\item Explain their thought process to answer the question
\item Share their answer
\item Read the scale out loud
\item Explain how they would improve the question or response scale
\end{itemize}

Recommendations were made to improve the questions after all sessions had been completed, and were integrated into the questionnaires for further testing during the pilot study (next section).

\subsubsection{Second round: Demographic, current use of AI and AI attitudes questions.}
The second round of cognitive testing was conducted with 8 full-time software engineers at Google, and focused on demographic, current use of AI, and AI attitudes questions. Study sessions repeated the procedure described above.

Major changes were made at this stage, including to questions that had been published and used by teams externally. For instance, some questions were double-barreled or vague, and participants struggled to answer them. The answer scales on other questions were not intuitive and therefore revised. Overall, we removed one question and added one, modified two questions significantly and edited one slightly, and rewrote multiple items and scales.

\subsection{Step 4: Identification of the benchmarking task}\label{idTask}

To benchmark our code AI features, we first needed to create a standardized task that research participants in the control and intervention groups would be required to complete. We settled on the following criteria:

\begin{itemize}
\item \textbf{Feasible within the context of the study}. We defined feasibility as a task that could be administered asynchronously and that could be completed within a time window for which we could reasonably expect participants to block off their calendar and focus on the task. In our case, this meant that the task would be expected to take fewer than 2.5 hours.
\item \textbf{High-fidelity and enterprise grade}. We defined ``high fidelity'' and ``enterprise grade'' as a task that most Google back-end engineers could reasonably be expected to complete as part of their daily work, that was complex enough that it would require writing code, building it, and testing it. This therefore excluded tasks that would be included in LeetCode \cite{leetcode}, for instance, or toy problems such as those in PyBench \cite{pybench}.
\end{itemize}

Once we had agreed on these criteria, the engineering team brainstormed multiple potential areas of work, using the GitHub Copilot JavaScript HTTP server task as a guide \cite{peng2023}. They settled on the creation of a Google-specific data logging server in C++, a dominant language at Google and in much of the software development industry. Other teams might choose a task that is more or less complex based on their developer population and on the end-to-end developer tasks that their features are expected to support.

\subsection{Step 5: Pilot to explore feasibility of the experimental task}\label{pilotFeasibility}
To ascertain the feasibility of our data logging task, we completed a pilot study with C++ developers with varying levels of experience with the creation of data logging systems at Google. In a lab context and under observation from the lead author (a researcher), participants were given a task description, which they were randomly assigned to complete with or without AI. Participants were given 1.5 hours to work on the problem while being recorded. The researcher conducted observations, took notes, and made adjustments to the instructions as the session progressed. An engineer was on call to address questions as needed and help adjust the task.

The first version of the task was not possible to complete within the 1.5 hour time frame, and the team started iterating to scope it more appropriately. We continued recruiting and iterating until the task was completed in under 1.5 hours \textit{and} without researcher intervention three times in a row, which took eight sessions in total. The final task description that was used in our experiment can be found in Appendix \ref{app:taskDescr}.

\section{Experimental Design}
\label{sec:researchDesign}

In the context of benchmarking complex tasks, randomized controlled human experiments such as the one in the present study have multiple benefits. They enable the use of a standardized task, something that is impossible at the ecosystem level. They enable the exploration of the role of personal attributes and attitudes. They enable comparison over time and, in our study design, across multiple features. More importantly, unlike observational studies or other qualitative research designs, the use of a control group enables causal analysis~\cite{Deaton2018}.

\begin{figure*}
\centering
\includegraphics[width=0.75\linewidth]{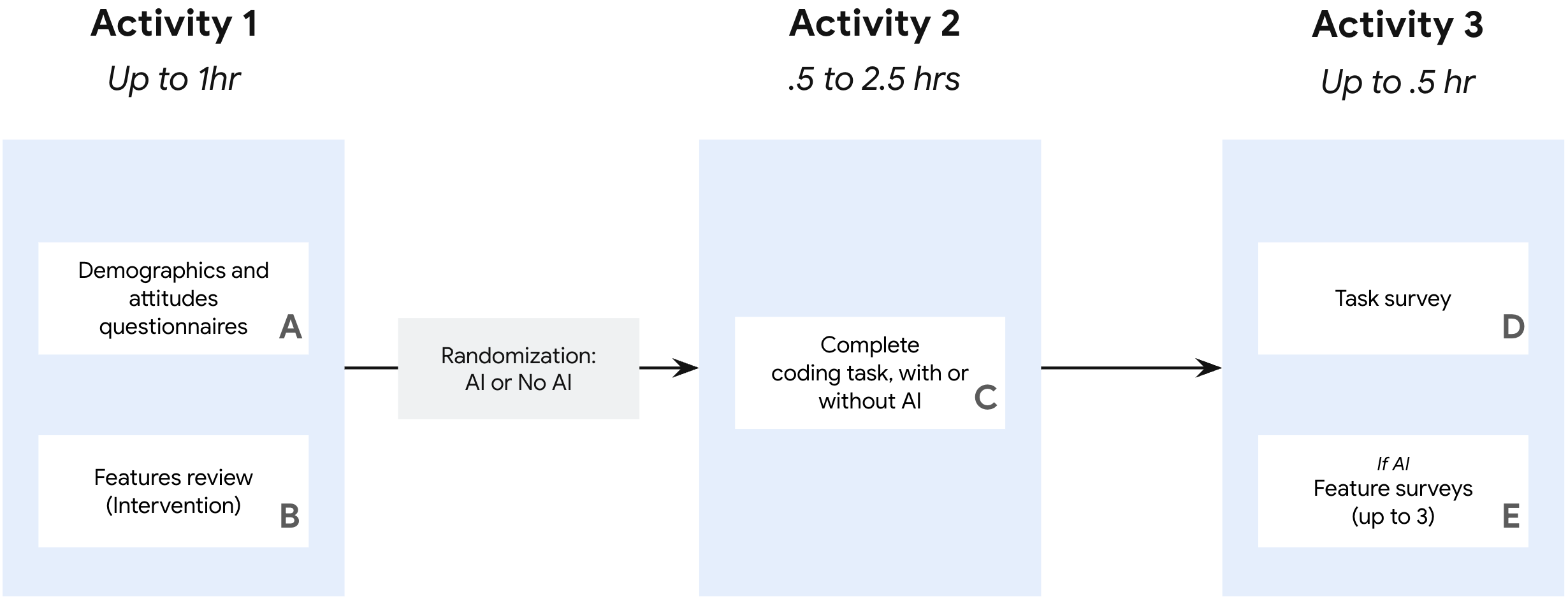}
\caption{Experimental design and components}
\label{fig:visualStudy}
\end{figure*}

A schematic overview of our between-subject human experimental design and its components can be found in Figure~\ref{fig:visualStudy}. We selected a between-subject design for multiple reasons. First, it would have been impossible to re-use the task in a within-subject design and make participants complete it with and without AI, because of unavoidable learning effects. Second, it was very difficult to design even one task that would be feasible by a broad-enough swath of developers with varied previous knowledge and skills. Designing two tasks with similar complexity was not feasible given the constraints on our task and on our time (see Section~\ref{idTask} above). Third and finally, we were asking developers for a large time commitment (a total of up to four hours over three activities); adding a second coding activity would not have been practically feasible.

The full study consisted of three activities, designed to total at most four hours of participants' time. All activities were self-administered (i.e., unmoderated); participants received instructions in their calendars before each task. Members of the research team were available to answer questions through email or direct message, but given extensive pre-testing we received few requests, and most of them about scheduling. The main outcome for the experiment was measured as time on task for both the control group (no access to AI features) and the experimental group (access to AI features).

Activity 1 consisted of two steps, as represented in Figure \ref{fig:visualStudy}:  the completion of a questionnaire that evaluated participants' experience and attitudes towards AI (Box A); and a review of the features that would be available to participants during the experiment (Box B). Materials provided to participants included links to documentation and feature demos. After participants had completed Activity 1, they were randomly assigned to either the control (no AI) or experimental condition (with AI) ahead of Activity 2.

We elected to use an intervention to teach developers about our features (see Box B) for three reasons. First, this intervention allowed us to obtain an estimate of the impact of AI on users who are actually familiar with the tools whose impact we were evaluating. Second, exposing developers to our AI features in production helped them better evaluate the impact of \textit{not} having access to AI features on their development work, were they to be assigned to complete the task without AI. Third and more pragmatically, the possibility to gain new knowledge and skills with AI tools gave developers another, non-pecuniary form of compensation for their time, since our ability to compensate participants is constrained by internal DX research policies. 

Activity 2 consisted in the coding task itself (Box C). Developers were asked to create a service to log messages sent by a fake new product, and to store the messages according to specific instructions. The AI-condition version of the task is detailed in Appendix~\ref{app:taskDescr}. Instructions for the task were identical regardless of randomization except for the setup instructions: participants in the control group were instructed to disable all AI features, while participants in the experimental group were instructed to enable them.

The final activity, Activity 3, included a task survey (Box D) and up to three feature surveys, depending on participants' use of the features during Activity 2 (Box E).
\section{Project impact and considerations for adoption of components}
\label{sec:impactnext}

In this case study, we have described how we developed components to benchmark our AI-enhanced code features in a way that enables comparison across features and over time. We have also shared our research design and our final benchmarking components. In doing so, we aim to enable the industry to move faster together on benchmarking the quality of AI features for code from a developer experience perspective. We want to encourage others to borrow our components to benchmark their own genAI for code products, so that as an industry we can better compare products.

Before discussing the considerations that teams should explore before embarking on their own benchmarking journey, we first summarize  the impact this project has had at Google.

\subsection{Impact}

The impact of our benchmark can be summarized along three lines, which align with the promise of robust DX benchmarks: (1) to help measure the value offered by products to customers, (2) to compare our offerings against our competitors', and (3) to track progress over time.

The first contribution of this benchmark was to provide an estimate of the return of investment (ROI) provided by our developer suite of products. This was very important for our leadership team, who wanted to be able to answer questions about the impact derived out of model training and AI feature development on the company's bottom line.

Second, our ROI estimate enabled ballpark comparisons between our product suite and similar estimates published about Microsoft GitHub Copilot's over the past year~\cite{peng2023,cui2024}. These comparisons are critical for business decisions to build versus buy, and therefore, helped explore the potential of future investments in genAI for coding use cases.

Third, with a single study we were able to evaluate the impact of three different features on time spent on task by developers \textit{and} compare our developers' sentiment towards these features. We were thus able to learn which features and what aspects of these features are working more or less well for our developers, both in objective (speed) and subjective (feature sentiment, see Table~\ref{tab:taskAIonly}) terms. This initial estimate of speed gained and perceptions will allow us to aim for systematic improvements in our features over time.

\subsection{Considerations for adoption}

Teams interested in benchmarking their products can elect different approaches when using the materials herein, depending on three main factors: whether the task they need to benchmark is complex; whether their products are code-based; and whether or not their user population is technical.

Teams who want to benchmark relatively simple tasks that leverage genAI should likely elect an approach that more closely resembles traditional usability testing, applied to their users' core journeys. This type of benchmarking will provide a first yardstick against which to measure progress over time. The usefulness of our components will therefore be limited, although the feature questionnaires might contain useful questions, and the overall attitudes towards AI questions could help better understand users.

In contexts where tasks are complex and the AI solution is based in code, using our components might enable a quick win. Not only will teams be able to move fast in creating their own benchmark by using already-tested instruments, but they will be able to compare their estimates with others across the industry---including Google's. For such a use case, minimal tweaking of the survey questions would be optimal, and adjustments to the task to align with the environment in which AI will be benchmarked will be ideal.

Likewise, in contexts where tasks are complex and the AI solution is also aimed at technical roles but is not based in production code (such as data science use cases, for instance), using our components might also accelerate benchmarking. Adjusting the task to fit the product use case, and editing survey questions to align with users' context will minimize the work required to benchmark products or features.

At last, teams who support complex genAI-enhanced products that are neither technical nor geared towards tech-savvy users might have to repeat part(s) of the process we describe here. Before setting off on this journey, however, they might want to consider how much of a priority this is for their product team, given the costs associated with such an approach. Specifically, the time and resources required to execute the approach described in this paper were significant, and spanned almost a full calendar year. Moreover, they might want to question how much statistical significance matters to them, since power calculations on previous experiments (including ours \cite{paradis2024rct}) suggest that more than 100 participants are necessary to obtain statistical significance with such a study. Finally, teams might want to evaluate the maturity of their logging infrastructure, for such a study is not likely to be very rigorous if time on task cannot be measured precisely and instead relies on self-report.
\section{Conclusion}
Despite the flaws and constraints on model benchmarking \cite{mcintosh2024}, the industry is moving forward with evaluating model quality, without systematic approaches to benchmarking \textit{product} quality. Our case study will hopefully spark conversations about how we might further flesh out what makes for good AI-enhanced code \textit{products} and then evaluate them, so that we might strike a better balance between model quality and the quality of the developer experience~\cite{AIsoftEng}. We hope that this contribution will help accelerate the rise of a countervailing perspective on what matters for the success of AI-enhanced code products, and enable cross-functional work that benefits users.

\begin{acks}
The authors wish to acknowledge the contributions of the following people: Don Eriko Anselmo, Paige Bailey, Satish Chandra, Rico Cruz, Tao Dong, Madhura Dudhgaonkar, Brett Durrett, Mona El Mahdy, Mike Giardina, Shivani Govil, Kate Grey, Joshua Katz, Min Kim, Angelo Luo, Quinn Madison, Dan Mcclary, Ryan McGarry, Vahid Meimand, Cody Miller, Kristof Molnar, Daye Nam, Nicole Ortiz, Sara Ortloff, Robin Savinar, Johann Scheidt, Niranjan Tulpule, Nan Zhang, and all study participants.
\end{acks}

\bibliographystyle{ACM-Reference-Format}
\bibliography{references}

 \begin{appendices}
\appendix
\section{Findings from our literature review of AI metrics}\label{app:litreview}

The tables below list the questions we considered for Activity 1, when participants' shared their \textit{use} of AI tools (Table \ref{tab:useAI}), \textit{openness} about AI tools (Table \ref{tab:opennessAI}), perception of the \textit{impact} of AI tools on their work (Table \ref{tab:impactAI}), \textit{trust in the accuracy} of AI tools (Table \ref{tab:trustaccAI}), and \textit{general attitudes} towards technology and AI (Table \ref{tab:genattAI}).

\begin{table}[ht!]
\caption{Questions about the \textit{use} of AI tools} \label{tab:useAI}
\begin{tabular}{|p{0.3\linewidth}|l|l|p{0.22\linewidth}|}
\hline
\textbf{Question} & \textbf{Source} & \textbf{Cog tested?} & \textbf{Rationale} \\ \hline
\RaggedRight Do you currently use AI tools in your development process at work? & \cite{stacko} & No & \RaggedRight Unlikely to be differentiating within study population \\ \hline
\RaggedRight What do you use AI for? & \cite{coderpad} & No & \RaggedRight Developers might not know what features are powered by AI \\\hline
\RaggedRight How frequently do you use the following features of the existing AI assistants for coding? & \cite{jetbrains} & No & \RaggedRight Developers might not know what features are powered by AI\\\hline
\RaggedRight In which parts of your development workflow are you currently using AI tools and which are you interested in using AI tools for over the next year? & \cite{stacko} & Yes & \RaggedRight Broad swath of workflow areas\\\hline
\end{tabular}
\end{table}

\begin{table*}[ht!]
\caption{Questions about \textit{openness} to AI tools} \label{tab:opennessAI}
\renewcommand{\arraystretch}{1.1}
\begin{tabular}{|p{0.45\linewidth}|p{0.07\linewidth}|p{0.1\linewidth}|p{0.25\linewidth}|}
\hline
\textbf{Question} & \textbf{Source} & \textbf{Cog tested?} & \textbf{Rationale} \\ \hline
\RaggedRight How favorable is your stance on using AI tools as part of your development workflow? & \cite{stacko} & No & \RaggedRight Biased towards the positive and unlikely to be differentiating within study population \\\hline
\RaggedRight How likely is it that you would delegate the following activities to an AI assistant (in an ideal world where the performance of an AI assistant is humanlike)? & \cite{jetbrains} & Yes & \RaggedRight Signals openness to AI delegation across workflow areas\\\hline
\RaggedRight Would you like to use more AI as part of your job? & \cite{coderpad} & No & \RaggedRight Biased towards the positive \\\hline
\RaggedRight Do you agree with the following statements? e.g. “I have security concerns about using AI generation services”, “I am ready to use cloud-based AI generation services for work tasks” & \cite{jetbrains} & No & \RaggedRight Biased towards the positive and many items are irrelevant in local context \\\hline
\end{tabular}
\end{table*}

\begin{table*}[ht!]
\caption{Questions about the \textit{impact} of AI tools} \label{tab:impactAI}
\renewcommand{\arraystretch}{1.1}
\begin{tabular}{|p{0.45\linewidth}|p{0.07\linewidth}|p{0.1\linewidth}|p{0.25\linewidth}|}
\hline
\textbf{Question} & \textbf{Source} & \textbf{Cog tested?} & \textbf{Rationale} \\ \hline
\RaggedRight For the AI tools you use as part of your development workflow, what are the MOST important benefits you are hoping to achieve? & \cite{stacko} & No & \RaggedRight Biased towards the positive, neither focused on experiences or attitudes \\\hline
\RaggedRight How similar or different is your current development workflow compared before you started using AI tools? & \cite{stacko} & Yes & \RaggedRight Straight\-forward assessment of impact, especially when followed by open-ended question \\\hline
\RaggedRight Do you think that AI-assisted tools will help reduce your workload? & \cite{coderpad} & No & \RaggedRight Biased towards the positive, hypothetical and future-oriented \\\hline
\end{tabular}
\end{table*}

\begin{table*}[ht!]
\caption{Questions about the \textit{trust in accuracy} of AI tools} \label{tab:trustaccAI}
\renewcommand{\arraystretch}{1.1}
\begin{tabular}{|p{0.45\linewidth}|p{0.07\linewidth}|p{0.1\linewidth}|p{0.25\linewidth}|}
\hline
\textbf{Question} & \textbf{Source} & \textbf{Cog tested?} & \textbf{Rationale} \\ \hline
\RaggedRight How much do you trust the accuracy of the output from AI tools as part of your development workflow? & \cite{stacko} & Yes & \RaggedRight Explicit definition of where trust should be evaluated \\\hline
\RaggedRight In the last 3 months, how much did you trust the quality of the output from AI powered tools as part of your development work? & \cite{dangeloEngSat} & Yes & \RaggedRight Same as above, plus time scoping \\\hline
\end{tabular}
\end{table*}

\begin{table*}[ht!]
\caption{Questions about \textit{general attitudes} towards technology and AI} \label{tab:genattAI}
\renewcommand{\arraystretch}{1.1}
\begin{tabular}{|p{0.45\linewidth}|p{0.07\linewidth}|p{0.1\linewidth}|p{0.25\linewidth}|}
\hline
\textbf{Question} & \textbf{Source} & \textbf{Cog tested?} & \textbf{Rationale} \\ \hline
\RaggedRight Thinking about how your workflow and process changes over time, how similar or different do you anticipate your workflow to be 1 year from now as a result of AI tools you are currently using? & \cite{stacko} & No & \RaggedRight Hypothetical and future-oriented\\\hline
\RaggedRight Which option most closely describes your current outlook on generative AI? & \cite{coderpad} & Yes & \RaggedRight Broad range of emotions\\\hline
\RaggedRight Affinity for technology scale (10 items) & \cite{edison2003} & No & \RaggedRight Unlikely to be differentiating within study population\\\hline
\RaggedRight If `technophobia' is defined as feeling discomfort about computers or any new technology, which of the following best describes you? & \cite{edison2003} & No & \RaggedRight Unlikely to be differentiating within study population\\\hline
\RaggedRight General attitudes towards Artificial Intelligence scale (20 items) & \cite{SCHEPMAN2020} & No & \RaggedRight Unlikely to be differentiating within study population \\\hline
\RaggedRight General attitudes towards Artificial Intelligence scale (8 items) & \cite{BERGDAHL2023} & No & \RaggedRight Unlikely to be differentiating within study population \\\hline
\RaggedRight Please tell me whether you think each of the following will mostly help or mostly harm people in this country in the next 20 years. If you don't have an opinion about this, please just say so. & \cite{neudert2020} & No & \RaggedRight Unlikely to be differentiating within study population \\\hline
\end{tabular}
\end{table*}

\cleardoublepage

\section{Demographics \& attitudes questionnaires} \label{app:demogAttQ}

\begin{table}[H]
\caption{Demographics questionnaire (Box A in Figure \ref{fig:visualStudy}): Questions, anchors, question type, and randomization}
\label{tab:demogQs}
\renewcommand{\arraystretch}{1.1}
\centering
\begin{tabular}{|p{0.34\linewidth}|p{0.3\linewidth}|p{0.13\linewidth}|p{0.12\linewidth}|}
\hline
\textbf{Question} & \textbf{Anchors} & \RaggedRight \textbf{Question type} & \textbf{Rand'n} \\ \hline
\RaggedRight What is your job title at Google? & \RaggedRight [List of engineering titles]; Other & \RaggedRight Single choice & None \\\hline
\RaggedRight  How long have you worked at Google? & \RaggedRight < 1 year; 1-4 years; 5-8 years; 9+ years; Prefer not to answer & \RaggedRight Single choice & None\\\hline
\RaggedRight  How many years of professional programming experience do you have? & \RaggedRight < 1 year; 1-4 years; 5-8 years; 9+ years; Prefer not to answer & \RaggedRight Single choice & None\\\hline
\RaggedRight  Over the last three months, how many hours have you spent on development tasks per day, on average? & \RaggedRight < 1 hour; 1-4 hours; 5-8 hours; 9-12 hours; 13+ hours; Prefer not to answer & \RaggedRight Single choice & None\\\hline
\RaggedRight  In the last three months, which languages have you used the most? Select up to 3. & \RaggedRight [List of languages in use at Google]; Other; Prefer not to answer & \RaggedRight Multiple choice; max 3 & None\\\hline
\RaggedRight  [From the three selected] Please select the programming language you have used most regularly in the last three months. & \RaggedRight [Piped from previous question] & \RaggedRight Single choice & None \\\hline
\RaggedRight  What IDE do you use most often in your work at Google? & \RaggedRight [List of IDEs in use internally at Google]; Other & \RaggedRight Single choice & None\\\hline
\RaggedRight  Where is the code you’ve worked on in the last 3 months for Google hosted? Select all that apply. & \RaggedRight [List of repositories available internally at Google]; Other & \RaggedRight Single choice & All but last \\\hline
\end{tabular}
\end{table}

\begin{table*}[hb!]
\caption{Attitudes and experience questionnaire (Box A in Figure \ref{fig:visualStudy}): Questions, anchors, question type, and randomization}
\renewcommand{\arraystretch}{1.2}
\begin{tabular}{|p{0.25\linewidth}|p{0.35\linewidth}|l|p{0.13\linewidth}|}
\hline
\textbf{Question} & \textbf{Anchors} & \textbf{Question type} & \textbf{Randomization} \\ \hline
In the last 3 months, have you done any of the following specialized types of development work within Google? Select all that apply. &  ML / AI: tool use or exploration; ML / AI: development, production support, or maintenance; ML / AI: research; None of the above; Prefer not to answer & Multiple choice & First three\\\hline
What is your current level of expertise with machine learning or AI? & I am a newcomer, starting to explore this skill; I am a beginner, starting to apply my skills; I am familiar, confident with my skills; I am an expert, people come to me for advice; I don’t have this skill; I don’t know what this is & Single choice & \RaggedRight Reverse, except for last two\\\hline
\RaggedRight Over the past three months, how often have you been using AI tools in the following parts of your development workflow? & \RaggedRight \textit{Rows}: Writing code; Adding comments to your code; Debugging; Code explanation; Learning about a codebase; Writing tests; Preparing your CLs; Reviewing code; Deployment; Monitoring; Project planning; Writing documents; Collaborating with teammates. \textit{Scale}: Every day; More than once a week; Once a week; Once a month; About once in the past 3 months; Never; I’m not sure. & Matrix & \RaggedRight Reverse, except for last\\\hline
\RaggedRight How different is your current development workflow today, compared to before you started using AI tools? & Extremely different; Very different; Moderately different; Slightly different; Not at all different; I don't use AI tools for development work; I'm not sure & Single choice & \RaggedRight Reverse, except for last two\\\hline
\RaggedRight What specifically is different about your current development workflow, compared to before you started using AI tools? & N/A & Open ended & N/A \\\hline
\RaggedRight What would you do if you had access to an AI assistant for the following activities (assuming that the assistant has high accuracy)? & \RaggedRight \textit{Rows}: Writing code; Adding comments to your code; Debugging; Code explanation; Learning about a codebase; Writing tests; Preparing your CLs; Reviewing code; Deployment; Monitoring; Project planning; Writing documents; Collaborating with teammates. \textit{Scale}: I would delegate to AI fully; I would supervise the AI loosely; I would supervise the AI closely; I would still do it myself; I'm not sure & Matrix & \RaggedRight Reverse, except last\\\hline
\RaggedRight Which option(s) most closely describe(s) your current outlook on generative AI and your career? Select all that apply. & Optimistic: I am excited for the opportunities and efficiencies that this technology can bring; Neutral: I think the technology can help improve my workflow but don’t anticipate a huge impact on my career otherwise; Skeptical: I am concerned about the ethics of implementation or potential for misuse; Worried: I am worried about the impact of the technology’s impact on my career or job prospects; I prefer not to answer & Multiple choice & \RaggedRight Reverse, except for last\\\hline
\RaggedRight How have AI-powered features in developer tools impacted your productivity in your development workflow? & \RaggedRight Very positively; Somewhat positively; No impact; Somewhat negatively; Very negatively; I'm not sure & Single choice & \RaggedRight Reverse, except for last\\\hline
\end{tabular}
\end{table*}

\clearpage

\section{Post-task and features surveys}\label{app:taskFeaturesSurvey}

\begin{table*}[hb!]
\caption{Post-task survey (all participants; Box D in Figure \ref{fig:visualStudy}): Questions, anchors, question type, and randomization} \label{tab:taskAll}
\renewcommand{\arraystretch}{1.2}
\centering
\begin{tabular}{|p{0.35\linewidth}|p{0.3\linewidth}|p{0.1\linewidth}|p{0.12\linewidth}|}
\hline
\textbf{Question} & \textbf{Anchors} & \RaggedRight \textbf{Question type} & \textbf{Rand'n} \\ \hline
\RaggedRight How familiar were you with the task you were asked to accomplish in the experiment? & \RaggedRight Extremely familiar; Very familiar; Moderately familiar; Slightly familiar; Not at all familiar; I’m not sure & \RaggedRight Single choice & \RaggedRight Reverse, except for last\\\hline
\RaggedRight How important are similar tasks (building logging systems) in your everyday development workflow? &  \RaggedRight Extremely important; Very important; Moderately important; Slightly important; Not at all important; I’m not sure & \RaggedRight Single choice & \RaggedRight Reverse, except for last\\\hline
\end{tabular}
\end{table*}

\begin{table*}[hb!]
\caption{Post-task survey (participants in the ``no-AI'' condition; Box D in Figure \ref{fig:visualStudy}): Questions, anchors, question type, and randomization} \label{tab:taskNoAIonly}
\renewcommand{\arraystretch}{1.1}
\begin{tabular}{|p{0.35\linewidth}|p{0.3\linewidth}|p{0.1\linewidth}|p{0.12\linewidth}|}
\hline
\textbf{Question} & \textbf{Anchors} & \RaggedRight \textbf{Question type} & \textbf{Rand'n} \\ \hline
\RaggedRight In this experiment, you were NOT allowed to use AI-powered features in [IDE]. How did this impact your productivity in this task? & \RaggedRight Very positively; Somewhat positively; No impact; Somewhat negatively; Very negatively; I’m not sure. & \RaggedRight Single choice & \RaggedRight Reverse, except for last \\\hline
\RaggedRight You answered that NOT using AI-powered features (positively impacted / negatively impacted / had no impact on) your productivity when completing the study task. Can you tell us why? & N/A  & \RaggedRight Open ended & \RaggedRight Question piping from previous question\\\hline
\end{tabular}
\end{table*}

\begin{table*}[hb!]
\caption{Post-task survey (participants in the ``AI'' condition; Box D in Figure \ref{fig:visualStudy}): Questions, anchors, question type, and randomization} \label{tab:taskAIonly}
\begin{tabular}{|p{0.32\linewidth}|p{0.25\linewidth}|p{0.15\linewidth}|p{0.15\linewidth}|}
\hline
\textbf{Question} & \textbf{Anchors} & \RaggedRight \textbf{Question type} & \textbf{Rand'n} \\ \hline
\RaggedRight Compared to how you normally accomplish similar tasks without AI, how satisfied are you with your developer tools? & \RaggedRight Much more satisfied; More satisfied; Neither more nor less satisfied; Less satisfied; Much less satisfied; I’m not sure & Single choice & \RaggedRight Reverse, except for last\\\hline
\RaggedRight How accurate were the AI predictions in this task? & \RaggedRight Extremely accurate; Very accurate; Moderately accurate; Slightly accurate; Not at all accurate; I’m not sure & Single choice & \RaggedRight Reverse, except for last\\\hline
\RaggedRight To what extent do you expect the AI to produce high quality code in this task? & \RaggedRight Very much; A lot; Somewhat; A little; Not at all; I’m not sure & Single choice & \RaggedRight Reverse, except for last\\\hline
\RaggedRight Compared with the level of control you were given over this task, what would be the ideal level of control for you in this context? & \RaggedRight I had much too much control; I had too much control; I had the right amount of control; I had too little control; I had much too little control; I’m not sure & Single choice & \RaggedRight Reverse, except for last \\\hline
\RaggedRight How have AI-powered features in developer tools impacted your productivity in this task? & \RaggedRight Very positively; Somewhat positively; No impact; Somewhat negatively; Very negatively; I’m not sure & Single choice & \RaggedRight Reverse, except for last\\\hline
\RaggedRight You answered that using AI-powered features (positively impacted / negatively impacted / had no impact on) your productivity when completing the study task. Can you tell us why? & N/A & Open ended & \RaggedRight Question piping from previous question\\\hline
\end{tabular}
\end{table*}

\begin{table*}[hb!]
\caption{Sample feature questionnaire (Box E in Figure \ref{fig:visualStudy}): Questions, anchors, question type, and randomization} \label{tab:featureSurveyQs}
\begin{tabular}{|p{0.3\linewidth}|p{0.3\linewidth}|l|p{0.13\linewidth}|}
\hline
\textbf{Question} & \textbf{Anchors} & \textbf{Question type} & \textbf{Rand'n} \\ \hline
\RaggedRight Please review the video in the description above. Did you use this feature [Feature name] as you completed the task today?& Yes; No; I'm not sure. (If yes: proceed with other questions.) & Single choice & First two \\\hline
\RaggedRight How accurate were the AI predictions in this feature [Feature name]? & Extremely accurate; Very accurate; Moderately accurate; Slightly accurate; Not at all accurate; I’m not sure & Single choice & \RaggedRight Reverse, except for last\\\hline
\RaggedRight To what extent do you expect the AI feature [Feature name] to produce high quality code?&  Very much; A lot; Somewhat; A little; Not at all; I’m not sure & Single choice & \RaggedRight Reverse, except for last \\\hline
\RaggedRight How easy or difficult was it to use this AI-enabled feature [Feature name]? & Very easy; Somewhat easy; Neither easy nor difficult; Somewhat difficult; Very difficult; I’m not sure & Single choice & \RaggedRight Reverse, except for last\\\hline
\RaggedRight Compared with the level of control you were given from this feature [Feature name], what would be the ideal level of control for you over this feature?& I had much too much control; I had too much control; I had the right amount of control; I had too little control; I had much too little control; I’m not sure & Single choice & \RaggedRight Reverse, except for last  \\\hline
\RaggedRight How helpful was the AI feature [Feature name] in completing this task? & Extremely helpful; Very helpful; Moderately helpful; Slightly helpful; Not at all helpful; I’m not sure & Single choice & \RaggedRight Reverse, except for last\\\hline
\RaggedRight How did this AI-enabled feature [Feature name] impact your ability to maintain a high level of focus? & Very positively; Somewhat positively; No impact; Somewhat negatively; Very negatively; I’m not sure & Single choice & \RaggedRight Reverse, except for last\\\hline
\RaggedRight How did this AI-enabled feature [Feature name] impact your productivity during this task? & Very positively; Somewhat positively; No impact; Somewhat negatively; Very negatively; I’m not sure & Single choice & \RaggedRight Reverse, except for last \\\hline
\RaggedRight How will this AI-enabled feature [Feature name] impact your productivity in the future? & Very positively; Somewhat positively; No impact; Somewhat negatively; Very negatively; I’m not sure & Single choice & \RaggedRight Reverse, except for last \\\hline
\RaggedRight How polished did this AI-enabled feature [Feature name] feel to you? & Extremely polished; Very polished; Moderately polished; Slightly polished; Not at all polished; I’m not sure & Single choice & \RaggedRight Reverse, except for last\\\hline
\end{tabular}
\end{table*}

\clearpage
\section{Task description} \label{app:taskDescr}
Here is the task description we used for the AI condition (Box C, Activity 2 in Figure \ref{fig:visualStudy}). Participants were given a change list (CL) with multiple pre-existing files, which they had to edit to be able to complete the task. Some elements have been redacted in brackets (i.e. [redacted]) to protect Google's intellectual property. The only difference between the ``AI'' and ``no AI'' task descriptions were instructions to either enable (with AI) or disable (no AI) our AI features.
\subsection*{Background}

In this study, you will be asked to complete a task using your internal tools. You will be assisted by our AI tools within your IDE. You can use your usual internal tools, as you usually would.

Earlier, you have been introduced to the features we are testing. As a reminder, the documentation for these features is here:
\begin{itemize}
\item Link to [Feature 1] documentation
\item Link to [Feature 2] documentation
\item Link to [Feature 3] documentation
\end{itemize}
Below is the information you need to get started with the task. Note that there are three steps.
\subsection*{Step 1. Enable AI features}
\begin{itemize}
\item Navigate to settings in [your IDE]. [Participants were given a screenshot, not shown here.] 
\item Search for [feature string], then enable all 5 options below. [Participants were given a screenshot, not shown here.]
\end{itemize}

\subsection*{Step 2. Get coding}

\subsubsection*{Goal}

We are the [fake product] site reliability engineering team. We are adding a new feature to the [fake product]: sharing updates about where you are walking. The [fake product] generates the messages. We need to implement a service that can receive log messages from [fake product] and store them into a file on [internal storage]. These logs should persist until the user begins a new walk, at which time the logs should be truncated. 

\begin{figure*}
\centering
\includegraphics[scale=.4]{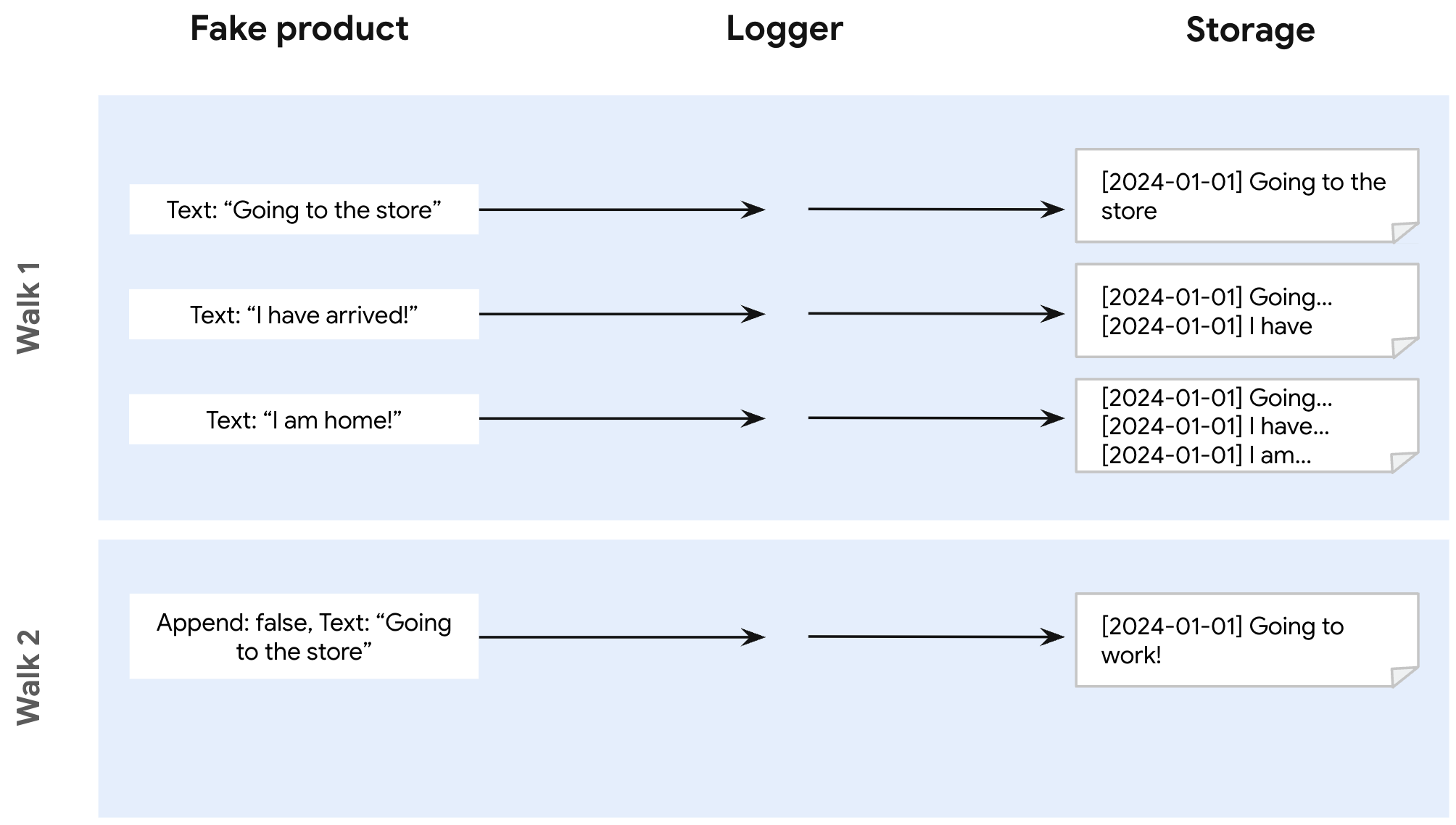}
\caption{Architecture provided to research participants}
\label{fig:architecture}
\end{figure*}

For a visual representation of the architecture, see Figure \ref{fig:architecture}. As you can see on this figure, each log message sent to the logger was appended to the file in [internal storage] until \texttt{append: false} was set on the request. Then the previous messages were discarded and the new message was all that existed in the file.

\subsubsection*{Setup} Create a new [source control] workspace in [your IDE] and call it \texttt{casestudy}. Please patch [a specific change list] in your IDE or terminal to get started. To simplify [access] for local development your application should write into [specific log file path]. For privacy and compliance reasons, each [fake product] should have its own file in [storage] where logs are written. If you have never used the sandbox in [internal storage] before, create the directory with your [personal ID] under [specific folder path].

\subsubsection*{Details}

\textbf{Server.} We have an existing [specific server architecture] server (not [another architecture]) into which we’d like you to implement this code.

The [specific server] should accept the following items:
\begin{itemize}
\item \texttt{application}. This is what the [fake product] team refers to as the user wearing the [fake product]. This will be used to name the file.
\item \texttt{event\_message}. This is the text the [fake product] has generated about the journey.
\item \texttt{append}: This should default to true and when set to false it should truncate/overwrite the log file.
\end{itemize}

\textbf{Log Format.} 
\begin{itemize}
\item An application name could have spaces or other special characters. Make sure that you replace them with underscores (i.e. "\_").

\item If we should add to the existing file, default is \texttt{append: true}. 
\item The server writes today's date, the event message, and a new line. For instance: "\texttt{2024-01-01 hello world \string\n}", where ``hello world'' is the message.
\item If the file does not exist and needs to be created, it needs to have [our data retention policy added to the file name].

\end{itemize}

\subsection*{Step 3. Wrap Up}

Make sure to update the existing [build files], [data structure files], and test files as needed. Build your code, then after you are done, test your work using the \texttt{log\_handler\_test.cc} file. You may need to edit the tests to work with your methods’ names.


 \end{appendices}

\end{document}